\documentclass[pdflatex,sn-mathphys-num]{sn-jnl}

\usepackage{graphicx}%
\usepackage{multirow}%
\usepackage{amsmath,amssymb,amsfonts}%
\usepackage{amsthm}%
\usepackage{mathrsfs}%
\usepackage[title]{appendix}%
\usepackage{xcolor}%
\usepackage{textcomp}%
\usepackage{manyfoot}%
\usepackage{booktabs}%
\usepackage{algorithm}%
\usepackage{algorithmicx}%
\usepackage{algpseudocode}%
\usepackage{listings}%
\usepackage[version=4]{mhchem}%

\theoremstyle{thmstyleone}%
\theoremstyle{thmstyletwo}%
\theoremstyle{thmstylethree}%

\begin{document}

%
%
%
%

\title[Chiral Intercalated 1T$'$-WS$_2$ Superlattices]{Chiral, Electronically Decoupled layers of 1T$'$-\ce{WS2} Topological Insulator via Neutral-Molecule Intercalation}

\author[1]{\fnm{Jiaze} \sur{Xie}}\email{jx3731@princeton.edu}

\author[1]{\fnm{Fatmagül} \sur{Katmer}}\email{fk6671@princeton.edu}

\author[1]{\fnm{Fang} \sur{Yuan}}\email{fy6@princeton.edu}

\author[1]{\fnm{Jaime M.} \sur{Moya}}\email{moya@princeton.edu}

\author[2]{\fnm{Guangming} \sur{Cheng}}\email{gcheng2@princeton.edu}

\author[1]{\fnm{Connor J.} \sur{Pollak}}\email{connorpollak@princeton.edu}

\author[3]{\fnm{Xiaoyu} \sur{Song}}\email{xs2509@columbia.edu}

\author[4]{\fnm{Nirmal}\sur{Roy}}\email{nirmalroy1378@gmail.com}

\author[4]{\fnm{Yakov}\sur{Bloch}}\email{yakovbloch@mail.tau.ac.il}

\author[4]{\fnm{Moshe Ben} \sur{Shalom}}\email{moshebs@tauex.tau.ac.il}

\author[5,6]{\fnm{Jennifer} \sur{Cano}}\email{jennifer.cano@stonybrook.edu}

\author*[1]{\fnm{Leslie M.} \sur{Schoop}}\email{lschoop@princeton.edu}

\affil*[1]{\orgdiv{Department of Chemistry}, \orgname{Princeton University}, \orgaddress{\street{Frick Chemistry Laboratory}, \city{Princeton}, \postcode{08544}, \state{NJ}, \country{USA}}}

\affil[2]{\orgdiv{Princeton Materials Institute}, \orgname{Princeton University}, \orgaddress{\street{Bowen Hall 70 Prospect Ave.}, \city{Princeton}, \postcode{08544}, \state{NJ}, \country{USA}}}

\affil[3]{\orgdiv{Department of Chemistry}, \orgname{Columbia University}, \orgaddress{\street{3000 Broadway}, \city{New York}, \postcode{10027}, \state{NY}, \country{USA}}}

\affil[4]{\orgdiv{School of Physics and Astronomy}, \orgname{Tel Aviv University}, \orgaddress{\street{ Shenkar Bld}, \city{Tel Aviv}, \postcode{69978}, \state{Tel Aviv District}, \country{Israel}}}

\affil[5]{\orgdiv{Department of Physics and Astronomy}, \orgname{Stony Brook University}, \orgaddress{\street{100 Nicolls Road}, \city{Stony Brook}, \postcode{11794}, \state{New York}, \country{USA}}}

\affil[6]{\orgdiv{Center for Computational Quantum Physics}, \orgname{Flatiron Institute}, \orgaddress{\street{162 Fifth Avenue}, \city{New York}, \postcode{10010}, \state{New York}, \country{USA}}}

\abstract{Monolayer 1T$'$-\ce{WS2} is predicted to be a two-dimensional topological insulator, but its intrinsic electronic properties are masked by strong interlayer coupling in its metallic and superconducting bulk parent phase, 2M-\ce{WS2}. Isolating monolayers by mechanical exfoliation is also hindered by this coupling, preventing experimental examination of monolayer's properties. Here we show that 2M-\ce{WS2} undergoes amine intercalation through a simple wet-chemical reaction, yielding superlattices in which the 1T$'$ layers are structurally preserved but electronically decoupled by neutral molecular spacers. Intercalation expands the interlayer spacing from 0.5 to 1-4 nm and reconstructs the stacking while preserving the intralayer 1T$'$ framework. Controlled (de)intercalation reversibly switches the system between a superconducting metal and an insulator with an activation gap matching that of the isolated monolayer. Density functional theory indicates that the electronically decoupled layers retain the nontrivial $\mathbb{Z}_2$ topology of the monolayer. Chiral amine intercalation further induces chiroptical activity in \ce{WS2} electronic transitions. Overall, the successful intercalation challenges the long-held view that group VIB dichalcogenides are inert toward neutral-molecule intercalation, and demonstrates molecular intercalation as a general chemical route for realizing monolayer-like topological-insulator physics and enabling chiral van der Waals superlattices in bulk single crystals.}


\maketitle

\section{Introduction}\label{sec1}

Since the isolation of graphene initiated the field of two-dimensional (2D) materials \cite{novoselov2004electric, novoselov2005two}, a wide range of atomically thin crystals have been identified, exhibiting exotic quantum phenomena that arise from reduced dimensionality. Among them, 2D topological insulators (TI), also known as quantum spin Hall (QSH) insulators, combine an insulating bulk with conducting helical edge states protected by time-reversal symmetry, and offer considerable potential for low-dissipation spintronics.\cite{kane2005quantum, bernevig2006quantum} Theory predicts that monolayer 1T$'$-\ce{MX2} (M = W, Mo; X = S, Se, Te) transition-metal dichalcogenides (TMDs), built from distorted-octahedral zig-zag metal chains, are promising 2D QSH insulators (Fig. 1a).\cite{qian2014quantum} Topological edge states have since been reported in monolayer 1T$'$-\ce{WTe2} \cite{tang2017quantum, wu2018observation} and 1T$'$-\ce{WSe2}.\cite{chen2018large} However, they have not yet been observed in 1T$'$-\ce{WS2}, which has the advantage of enhanced air stability compared to tellurides and thermal robustness among metastable phases.\cite{lai2021metastable, liu2024high}

\begin{figure}[h]
\centering
\includegraphics[width=1\textwidth]{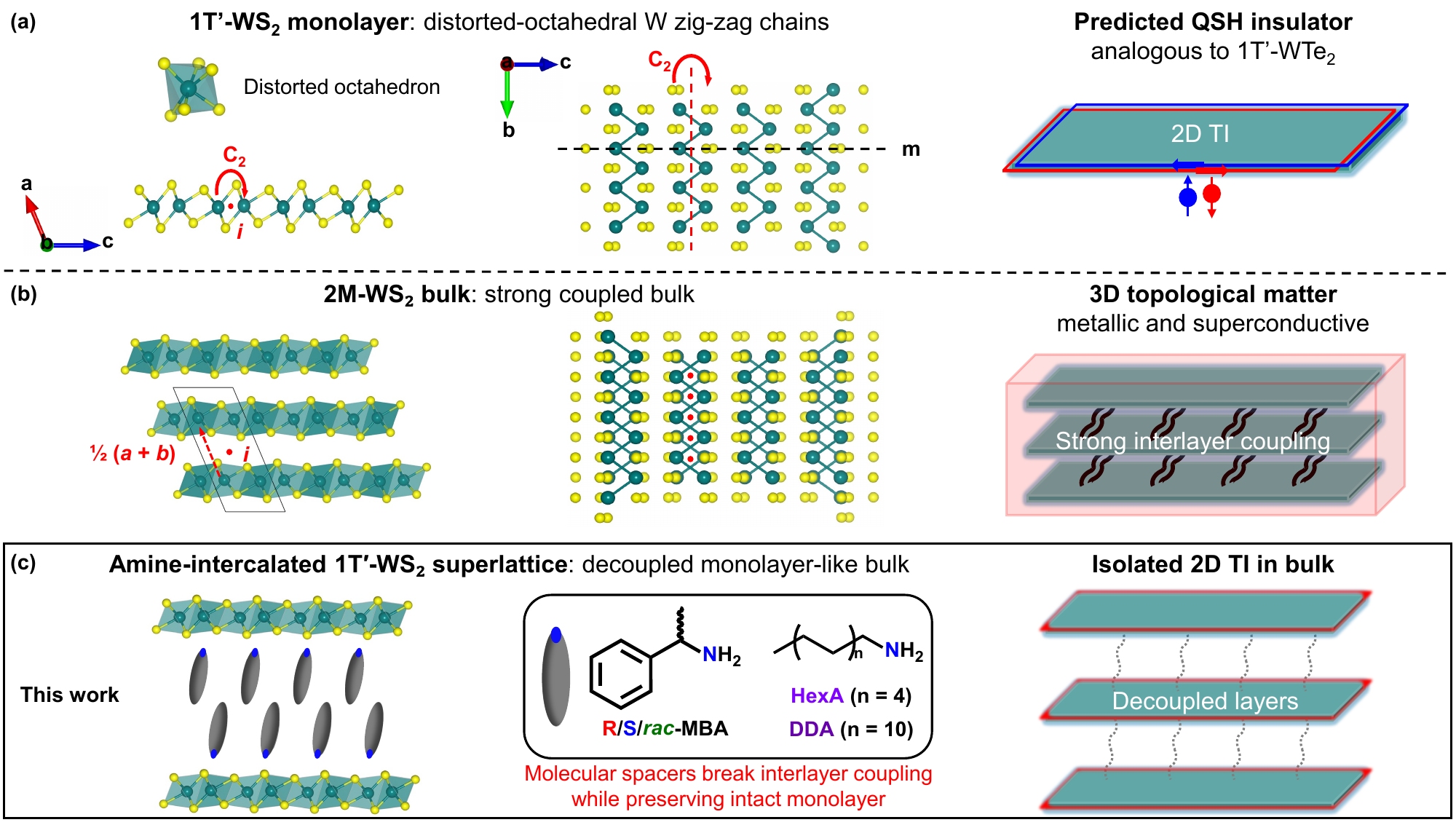}
\caption{\textbf{Crystal structures and topological hierarchy of \ce{WS2}.} 1T$'$-\ce{WS2} monolayer (a), 2M-\ce{WS2} bulk (b), and amine-intercalated \ce{WS2} superlattices (c). The 1T$'$ monolayer is predicted to be a 2D topological insulator (TI), namely a quantum spin Hall (QSH) insulator, whereas the bilayer-based 2M bulk phase, characterized by strong interlayer coupling, experimentally shows a 3D electronic state that is metallic and superconducting. In this work, neutral amine intercalation suppresses the interlayer coupling, decoupling the stacked 1T$'$ layers and restoring monolayer-like electronic behavior within a translational superlattice. W and S atoms are represented by cyan and yellow spheres, respectively. The amine intercalants are R/S/\textit{rac}-methylbenzylamine, \textit{n}-hexylamine, and \textit{n}-dodecylamine.}\label{fig1}
\end{figure}

Although monolayers across the 1T$'$-\ce{MX2} family share similar in-plane structures, their bulk parent phases exhibit fundamentally different stacking patterns of 1T$'$ layers, including the bulk 1T$'$, 1T$_d$, and 2M polymorphs associated with distinct interlayer couplings.\cite{qi2016superconductivity, xu2023topology} For instance, bulk 1T$_d$-\ce{WTe2} adopts a non-centrosymmetric orthorhombic stacking of relatively weakly coupled 1T$'$ layers (space group \textit{Pmn$2_1$}),\cite{brown1966crystal} whereas 2M-\ce{WS2}, the bulk phase of 1T$'$-\ce{WS2}, crystallizes in a base-centered monoclinic structure (space group \textit{C2/m}) with two 1T$'$ layers per unit cell.\cite{fang2019discovery} Adjacent layers are related by an (\textit{a}+\textit{b})/2 translation, where \textit{a} and \textit{b} denote lattice vectors, but remain tightly stacked, forming a centrosymmetric bilayer motif (Fig.~\ref{fig1}b). Angle-resolved photoemission spectroscopy (ARPES) and first-principles calculations both reveal pronounced interlayer coupling in 2M-\ce{WS2}, giving rise to significant 3D electronic dispersion rather than quasi-2D behavior.\cite{xu2023topology} Consistently, bulk 2M-\ce{WS2} stays metallic and becomes superconducting at low temperature even in samples as thin as 4 nm.\cite{wang2020nodeless, zhang2023spin, hossain2025tunable} Because of limitations in crystal quality and exfoliation, reliable transport on high-quality 1T$'$-\ce{WS2} monolayers remains elusive, and the intrinsic insulating nature of the monolayer has not been established experimentally, apart from photoemission evidence of a narrow-gap state in mixed 2H/1T$'$ nanosheets.\cite{pierucci2019evidence} Rather than pursuing cleaner monolayers with physical exfoliation, we introduce a chemical route: using neutral-molecule intercalation as an internal spacer to suppress interlayer coupling and recover monolayer-like electronic behavior within bulk crystals.

Neutral-molecule intercalation has a long history. Since the 1970s, \ce{TaS2} has been known to spontaneously intercalate Lewis-basic molecules, forming superlattices of intact TMD layers separated by insulating molecular spacers.\cite{gamble1971some} This chemistry enabled the early discovery of quasi-2D superconductivity prior to the development of mechanical exfoliation,\cite{gamble1970superconductivity, gamble1971intercalation} establishing a chemical route to electronically decoupled layered systems. However, such intercalation chemistry has largely been confined to metallic group IVB/VB TMDs, whereas group VIB compounds such as 2H-\ce{MoS2} and 2H-\ce{WS2} are generally considered inert toward neutral amines.\cite{rao1979intercalation} More recently, some amine-intercalated 1T$'$-\ce{MoS2} nanoparticles derived from \ce{(NH4)2MoS4} precursors have been reported for chiral transport and electrocatalytic applications,\cite{bian2022hybrid} hinting that metastable group VIB systems may exhibit a broader intercalation chemistry than previously recognized. The metastable 2M-\ce{WS2}, recently isolated by wet-chemical synthesis,\cite{fang2019discovery, song2023acid} is metallic and therefore a plausible host, yet its intercalation chemistry remains unexplored.

In this work, we intercalate 2M-\ce{WS2} with a series of neutral amines, namely R-, S-, and \textit{rac}-methylbenzylamine (R/S/\textit{rac}-MBA; \textit{rac} denotes the racemate), \textit{n}-hexylamine (HexA), and \textit{n}-dodecylamine (DDA), forming superlattices of electronically decoupled yet chemically intact 1T$'$-\ce{WS2} layers (Fig.~\ref{fig1}c). The intercalated phases are insulating, consistent with the ground state predicted for the monolayer, and intercalation with chiral amines imprints chiroptical activity onto the \ce{WS2} layers, providing a chemical handle for tuning electronic chirality. More broadly, these results demonstrate how wet-chemical intercalation can engineer van der Waals (vdW) heterostructures with tunable properties through host-guest interfacial proximity effects.\cite{liu2025precision}

\section{Results}\label{sec2}

\subsection{Synthesis and Chemical Characterization}\label{subsec2.1}

\begin{figure}[h]
\centering
\includegraphics[width=1\textwidth]{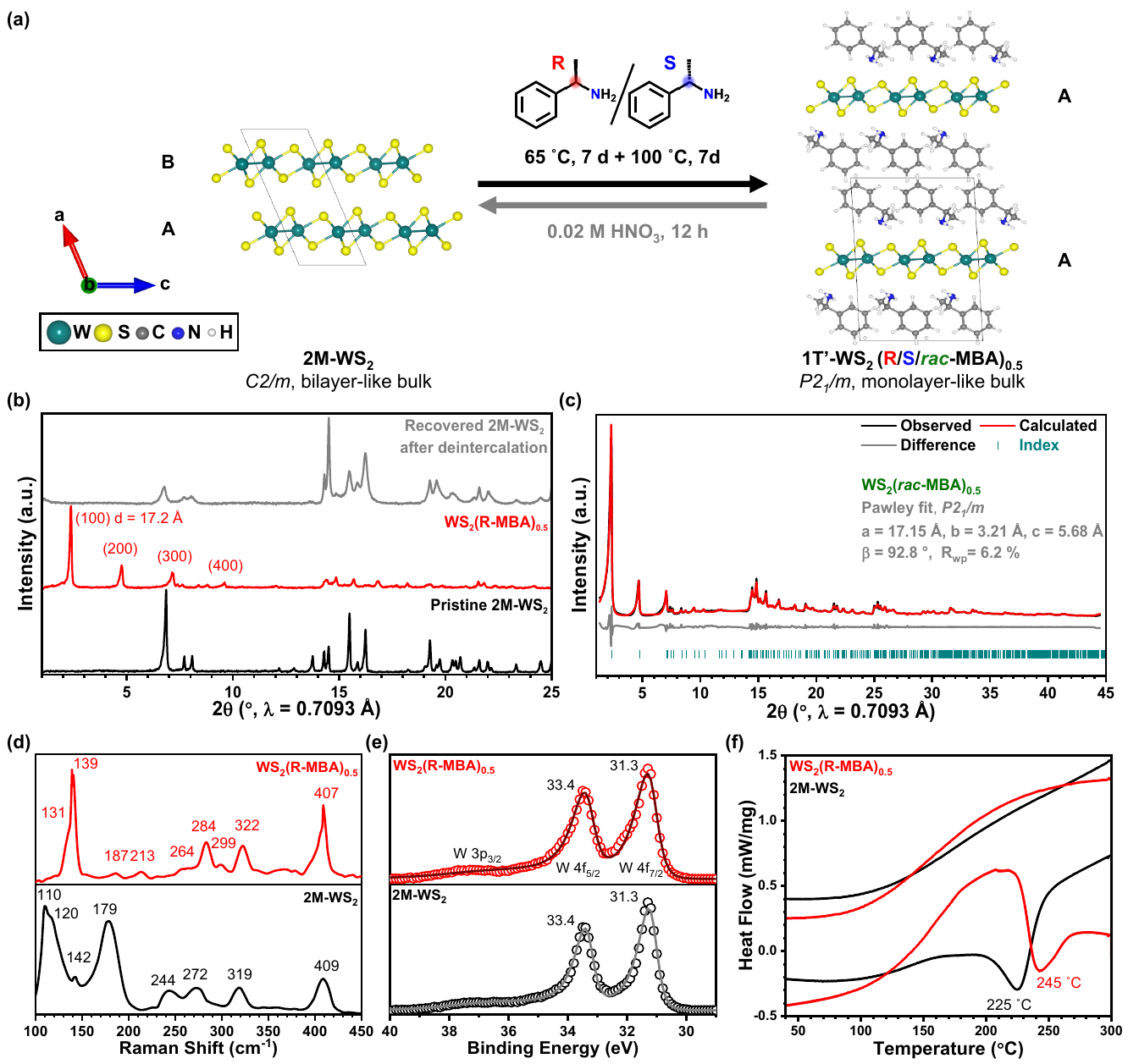}
\caption{\textbf{Synthesis and structural and chemical characterization of amine-intercalated \ce{WS2}, exemplified by R-MBA.} Schematic of reversible (de)intercalation in 2M-\ce{WS2} using MBA (a), with \textit{ex situ} PXRD tracking the structural evolution (b). Pawley refinement of \ce{WS2}(\textit{rac}-MBA)$_{0.5}$ confirms the intercalated phase (c). Raman (d), W 4f XPS (e), and DSC (f) collectively evidence the intercalation-induced structural and electronic changes. Full characterization of all amine-intercalated phases is provided in Extended Data Fig. 2.}\label{fig2}
\end{figure}

\ce{WS2}-amine hybrids were obtained by soaking 2M-\ce{WS2} crystals in anhydrous amines under heating for two weeks (Fig.~\ref{fig2}a). The \ce{WS2} stoichiometry before and after intercalation was determined by inductively coupled plasma optical emission spectroscopy (ICP-OES, Table S1) and corroborated by scanning electron microscopy and scanning-transmission electron microscopy with energy-dispersive X-ray spectroscopy (SEM/STEM-EDS, Fig. S12 to S18). Thermogravimetric analysis (TGA, Extended Data Fig. 1) indicates amine contents of approximately 0.5 molecules per W atom for all intercalates except HexA (0.3, see Methods), consistent with nearly complete intercalation.\cite{rao1979intercalation} Correspondingly, the low-angle (h00) reflections in powder X-ray diffraction (PXRD) reveal substantially expanded interlayer spacings along the \textit{a} axis (Fig.~\ref{fig2}b; for example, from 5.9 to 17.2~\AA\ for MBA), matching the expansion expected for an amine-bilayer arrangement\cite{jeong2015tandem} and supporting the formation of fully intercalated phases.\cite{powell1993intercalation}

X-ray photoelectron spectroscopy (XPS) shows no significant shifts in the W and S core levels, excluding substantial reductive charge transfer from the amines to \ce{WS2} during intercalation (Fig.~\ref{fig2}e, S8, Extended Data Fig. 2c). Differential scanning calorimetry (DSC) reveals the characteristic 1T$'$/2M-to-2H phase transition (Fig.~\ref{fig2}f, S21, Extended Data Fig. 2d),\cite{song2023acid} confirming that the metastable 1T$'$-\ce{WS2} framework is preserved. The elevated transition temperatures of the intercalated phases point to enhanced thermal stability, consistent with earlier work on mono- and few-layer 1T$'$-\ce{WS2},\cite{liu2024high} and indicate that the mild wet-chemical treatment does not introduce substantial defects, which would otherwise lower the transition temperature.\cite{song2023acid} Raman spectra retain the characteristic modes of 1T$'$/2M-\ce{WS2} but show several peak shifts, including a pronounced red shift of the mode near 120 cm$^{-1}$ (Fig.~\ref{fig2}d, Extended Data Fig. 2b), as expected for a change from bulk-like to monolayer-like behavior.\cite{liu2024high, li20241t} PXRD monitoring further shows that acid-assisted deintercalation fully restores the parent 2M-\ce{WS2} phase, with no detectable 2H or other impurity phases (Fig.~\ref{fig2}b), demonstrating reversible (de)intercalation without framework decomposition, analogous to the behavior reported for 2H-\ce{TaS2}(pyridine)$_{0.5}$.\cite{gamble1971some} Finally, Infrared spectra show only minor red shifts of guest-molecule vibrational modes upon intercalation, typically less than 50 cm$^{-1}$ (Fig.~S3). Neither the large frequency shifts (100-200 cm$^{-1}$) expected for protonation\cite{socrates2004infrared} nor the Fano-type line-shape distortions characteristic of covalent bonding\cite{zhou2023hybrid} are observed, indicating that the guest molecules remain electronically intact within the intercalated structure. 

Together, these results establish a neutral-molecule intercalation process analogous to that long known for group IVB/VB TMDs,\cite{schollhorn1977demonstration} yielding structurally intact yet electronically decoupled 1T$'$-\ce{WS2} superlattices. The mild, long-term solvothermal conditions produce highly crystalline, fully intercalated single phases, for which Pawley and Rietveld refinements support the proposed structural models (Fig.~\ref{fig2}c, S7), while high-resolution STEM (HRSTEM) will directly resolve the structural evolution at the atomic scale.

\subsection{Structural Characterization at Atomic Resolution}\label{subsec2.2}

\begin{figure}[h]
\centering
\includegraphics[width=1\textwidth]{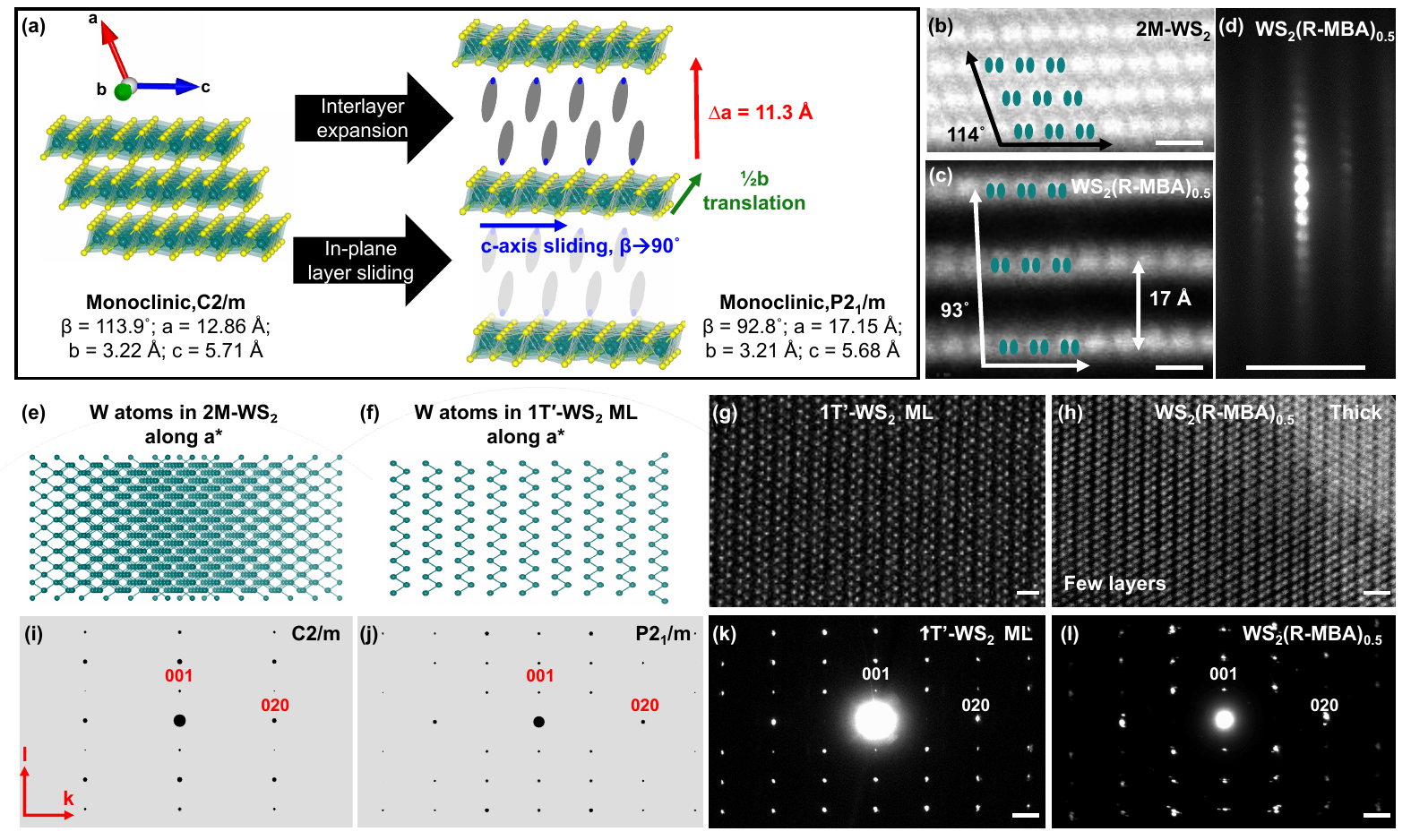}
\caption{\textbf{Structural evolution of amine-intercalated \ce{WS2} at atomic resolution, exemplified by R-MBA.} Schematic of the structural evolution associated with the lattice and symmetry changes (a). Cross-sectional HRSTEM (along the \textit{b}-axis) of 2M-\ce{WS2} (b) and \ce{WS2}(R-MBA)$_{0.5}$ (c), together with cross-sectional electron diffraction of \ce{WS2}(R-MBA)$_{0.5}$ (d). Simulated in-plane W atomic lattices and corresponding HRSTEM images of the 1T$'$-\ce{WS2} monolayer (ML, e, g) and \ce{WS2}(R-MBA)$_{0.5}$ (f, h) along the \textit{a}$^*$ projection. Simulated in-plane electron diffraction patterns of 2M-\ce{WS2} (i) and 1T$'$-\ce{WS2} (j), alongside experimental in-plane diffraction of the 1T$'$-\ce{WS2} monolayer (k) and \ce{WS2}(R-MBA)$_{0.5}$ (l), all viewed along the \textit{a}$^*$ direction. Scale bars represent 1 nm in the HRSTEM panels and 5 nm$^{-1}$ in the electron diffraction panels. The systematic absences in the 0kl plane for \textit{C2/m} are k = 2n+1, arising from the C-centering condition. In contrast, for \textit{P2$_1$/m}, the corresponding extinction condition is limited to 0k0 reflections with k = 2n+1, resulting from the \textit{2$_1$} screw axis along the \textit{b} direction.}\label{fig3}
\end{figure}

Despite the long history of molecular intercalation, only recently have amine-intercalated TMDs crystals such as 1T-\ce{TiS2},\cite{jeong2015tandem, bian2023chiral} and 2H-\ce{TaS2}\cite{qian2022chiral} been examined by HRSTEM. Since these hosts are trigonal or hexagonal, previous studies have focused mainly on interlayer expansion. For monoclinic 2M-\ce{WS2}, we instead find a much larger structural reconstruction upon intercalation, in which sliding of individual layers does not necessarily introduce disorder but effectively resets the interlayer symmetry (Fig.~\ref{fig3}a).

Cross-sectional high-angle annular dark-field STEM (HAADF-STEM) images taken along the \textit{b}-axis clearly resolve the ordered W dimers of the zig-zag W chains in the bc plane, for both pristine and intercalated phases, confirming that the intralayer 1T$'$-\ce{WS2} framework remains intact at atomic resolution (Fig.~\ref{fig3}b, c). The images also show expanded interlayer spacings of approximately 17~\AA\ along the \textit{a}-axis, consistent with electron diffraction (Fig.~\ref{fig3}d) and PXRD (Fig.~\ref{fig2}b). More importantly, the stacking changes markedly upon intercalation: the monoclinic $\beta$ angle decreases from $114^\circ$ in pristine 2M-\ce{WS2} to nearly orthogonal ($93^\circ$) in the intercalated phases (Fig.~\ref{fig3}b, c), indicating substantial interlayer shifts along the \textit{c}-axis. The same behavior is seen, even more clearly, in \ce{WS2}(HexA)$_{0.3}$, despite different interlayer expansion (Fig.~S20).

The stacking rearrangement also has a direct, visible consequence in plan-view imaging. In pristine 2M-\ce{WS2} with \textit{C2/m} symmetry, AB stacking offsets the zig-zag W chains of neighboring layers such that, along the \textit{a}$^*$ projection (perpendicular to the bc plane), chains from successive layers overlap rather than appearing as separated features (Fig.~\ref{fig3}e, S19). By contrast, in an isolated 1T$'$ layer (\textit{P2$_1$/m}, Fig.~\ref{fig3}f), the W chains project as well-separated parallel lines. The intercalated superlattices exhibit this monolayer-like pattern of clearly separated parallel chains across the entire flake and even in multilayer regions (Fig.~\ref{fig3}g,h),\cite{song2023synthesis} suggesting that layers are aligned directly above one another. This monolayer-like bulk structure indicates substantial layer shifts along both the \textit{b}- and \textit{c}-axes, together with a pronounced reconstruction of the interlayer registry (Fig.~\ref{fig3}a).

The symmetry lowering from \textit{C2/m} to \textit{P2$_1$/m} suggested by PXRD refinements is also confirmed by in-plane electron diffraction. Diffraction spots that are forbidden in \textit{C2/m} (odd $k$, Fig.~\ref{fig3}i) but allowed in \textit{P2$_1$/m} (where only 0$k$0 reflections with odd $k$ are forbidden, Fig.~\ref{fig3}j) emerge in the intercalated phases (Fig.~\ref{fig3}l), closely matching those of the monolayer (Fig.~\ref{fig3}k). We note that \textit{P2$_1$/m} is also the space group of monoclinic 1T$'$-\ce{MoTe2},\cite{puotinen1961crystal} a prototypical weakly coupled 1T$'$ layered system. \cite{xu2023topology}

Overall, molecular intercalation reconstructs the interlayer stacking while preserving the intralayer 1T$'$-\ce{WS2} framework. The resulting monoclinic-to-nearly-orthorhombic transition reflects a strong reduction of the interlayer coupling present in pristine 2M-\ce{WS2}, and the formation of weakly interacting \ce{WS2}-molecule units that favor near-parallel layer alignment. Because the \ce{WS2} layers stay intact, these materials let us probe monolayer-like physical properties within bulk single crystals.

\subsection{Physical Characterization}\label{subsec2.3}

\begin{figure}[h]
\centering
\includegraphics[width=0.9\textwidth]{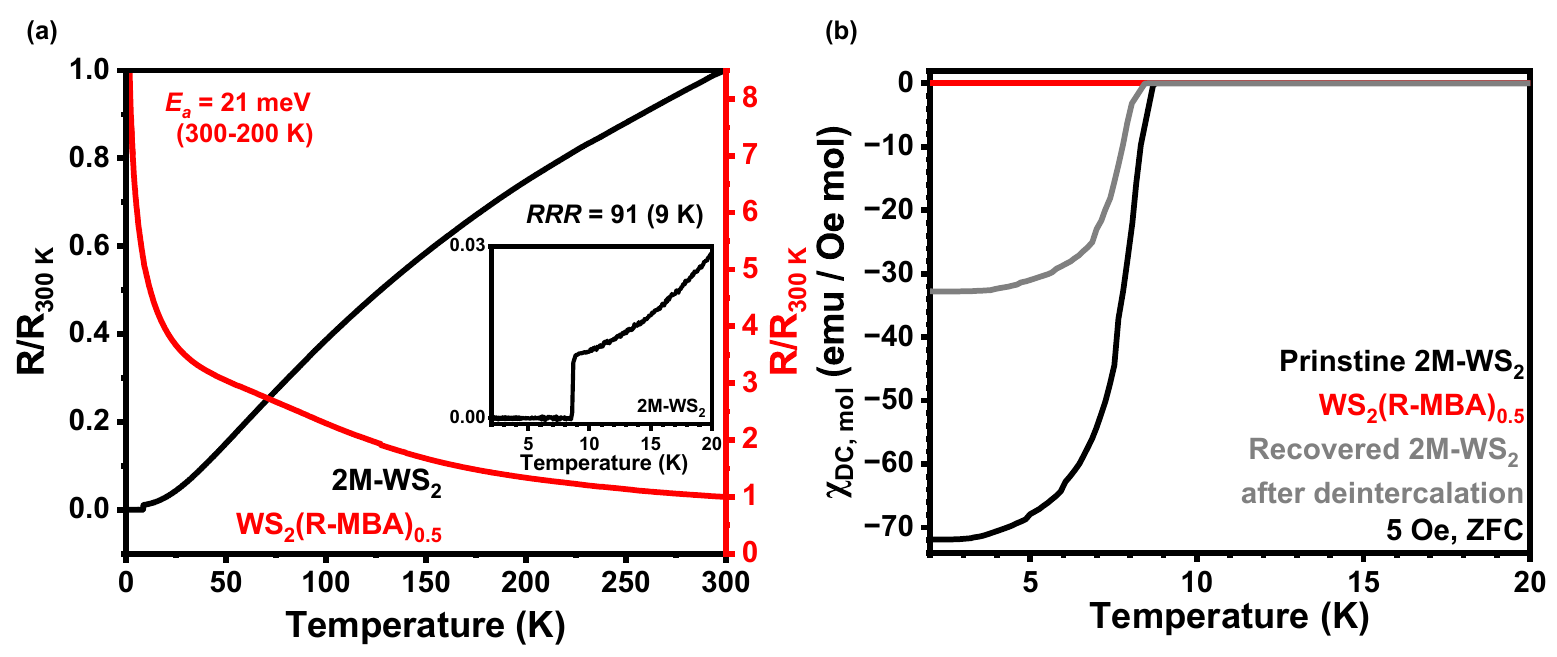}
\caption{\textbf{Physical properties before and after amine intercalation in \ce{WS2}, exemplified by R-MBA.} Temperature-dependent resistance of 2M-\ce{WS2} and \ce{WS2}(R-MBA)$_{0.5}$ crystals (a); the inset highlights the low-temperature superconductivity in 2M-\ce{WS2}. Magnetic susceptibilities of 2M-\ce{WS2}, \ce{WS2}(R-MBA)$_{0.5}$, and restored 2M-\ce{WS2} after deintercalation, highlighting the changes in diamagnetic response associated with superconductivity (b). Full characterization of all amine-intercalated phases is provided in Extended Data Fig. 3. Samples were zero-field cooled (ZFC) and subsequently measured under an applied magnetic field of 5 Oe. Field-cooled (FC) results are included in Fig.~S24.}\label{fig4}
\end{figure}

Four-probe transport was measured on single crystals of pristine 2M-\ce{WS2} and intercalated \ce{WS2} superlattices (Fig.~\ref{fig4}a, Extended Data Fig. 3a). Pristine 2M-\ce{WS2} is metallic, with resistance decreasing upon cooling and dropping to zero at 8.7 K as superconductivity emerges. Its residual resistance ratio (\textit{RRR}, $R_{300~\mathrm{K}}/R_{9~\mathrm{K}}$) reaches 91, confirming the high crystalline quality of the parent phase. In contrast, all amine-intercalated samples are insulating (d$R$/d$T<0$), with resistance increasing upon cooling and no superconducting transition observed. The insulating states exhibit activation energies of 20-50 meV ($E_a$, Fig.~S23), broadly consistent with the calculated band gap of monolayer 1T$'$-\ce{WS2} ($\sim$0.1 eV, corresponding to roughly 2$E_a$, Fig.~S25).\cite{qian2014quantum}

Magnetic susceptibility confirms this transition. Pristine 2M-\ce{WS2} shows a superconducting transition at 8.7 K in an applied field of 5 Oe, whereas all fully intercalated superlattices are nonmagnetic with no detectable superconductivity (Fig.~\ref{fig4}b, Extended Data Fig. 3b). Importantly, superconductivity is recovered after acid-assisted de-intercalation, demonstrating that the metal-to-insulator transition is reversible and not caused by permanent defects or chemical decomposition. Consistently, partially intercalated and de-intercalated samples that contain mixed pristine and intercalated phases also retain superconductivity (Extended Data Fig. 4, 5), indicating that superconductivity does not require perfect bulk 2M-\ce{WS2} stacking and can survive in regions of only a few imperfectly stacked 1T$'$-\ce{WS2} layers, in line with earlier reports of superconductivity in films restacked from 1T$'$-\ce{WS2} monolayer inks.\cite{song2023synthesis} Therefore, the insulating behavior of the fully intercalated superlattices most likely arises from the increased interlayer separation, which electronically isolates the 1T$'$-\ce{WS2} layers and suppresses interlayer coupling. Molecular intercalation thus drives an electronic crossover from a metallic superconducting state to an electronically decoupled, monolayer-like insulating state.

Notably, the reversible behavior we observe differs from the recently reported amine-intercalated \ce{TiSe2} system,\cite{song2025transitory} where sacrificial intercalants thermally decompose, irreversibly resulting in a new superconducting phase after deintercalation. This behavior is plausibly linked to the chemical sensitivity of \ce{TiSe2},\cite{hildebrand2014doping, sun2017suppression, moya2019effect} in which decomposition and side reactions with amines can accompany intercalation. By contrast, 2M-\ce{WS2} is exceptionally robust: it can be synthesized under strongly oxidative acidic conditions with chromate in air\cite{fang2019discovery, lai2021metastable, song2023acid} and has been used as an electrode for aqueous potassium-ion storage.\cite{mao2019k} Consistent with this robustness, the 2M-\ce{WS2} intercalation reactions maintain colorless solutions throughout, in contrast to the brown coloration observed for the \ce{TiSe2} system, indicating the absence of comparable degradation pathways. Thus, this framework stability supports the reversible changes in physical properties arising primarily from intercalation-induced stacking rearrangements rather than irreversible chemical transformation.

Together, these results show that molecular intercalation suppresses the metallic state of bulk 2M-\ce{WS2}, and that the resulting insulating superlattices match the insulating ground state predicted for isolated monolayer 1T$'$-\ce{WS2}.\cite{qian2014quantum, zhang2023spin} The chemical robustness of this system makes amine-intercalated \ce{WS2} superlattices an attractive platform for probing intrinsic electronic states without complications from environmental degradation.

\subsection{Theoretical Calculations}\label{subsec2.4}

\begin{figure}[H]
\centering
\includegraphics[width=0.9\textwidth]{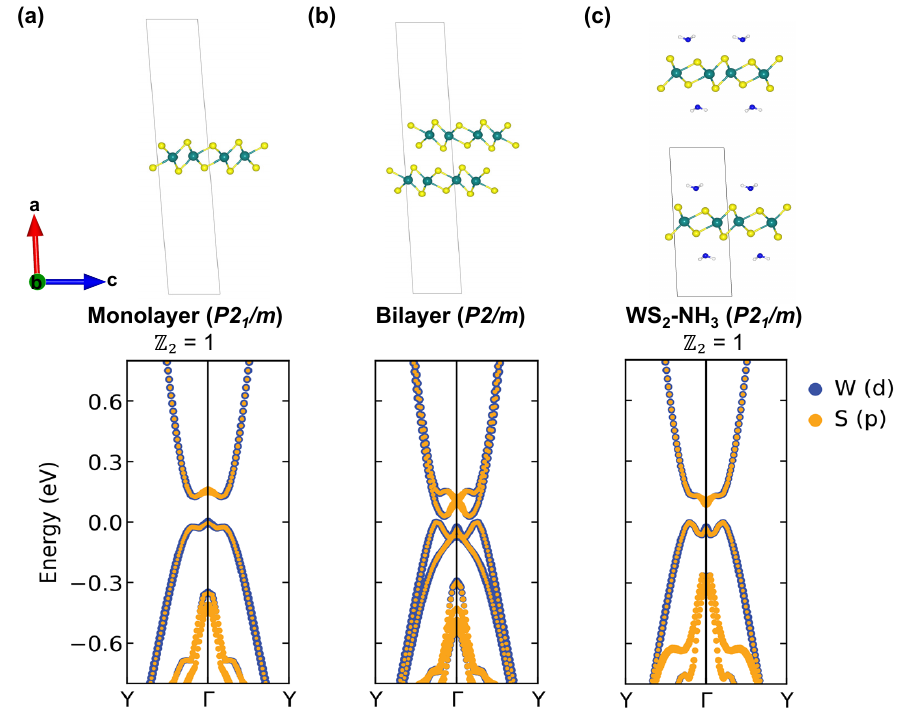}
\caption{\textbf{Topology is recovered upon intercalation.} Crystal structures and corresponding orbital-projected band structures along the $\Gamma$-Y path of the monolayer (a), bilayer (b), and 1T$'$-\ce{WS2}-\ce{NH3} superlattice (c). Topological analysis indicates that the \ce{WS2}-\ce{NH3} superlattice retains the nontrivial topology of the monolayer, characterized by $\mathbb{Z}_2 = 1$. Band structures along the full Brillouin-zone paths are provided in Extended Data Fig.~6.}\label{fig_dft}
\end{figure}

To understand the electronic structure of the superlattices, we carried out density functional theory (DFT) calculations on monolayer, bilayer, and intercalated 1T$'$-\ce{WS2}. Modeling bulky chiral amines requires large supercells, leading to significant computational costs; more importantly, low-symmetry intercalants lower the space group to \textit{P1}, eliminating the inversion center required for evaluating the topological indices. We therefore constructed a \ce{WS2}-\ce{NH3} model in which \ce{NH3} molecules are placed near the \ce{WS2} layers while the interlayer spacing is fixed to the experimental value ($\sim$17~\AA) refined from \ce{WS2}(\textit{rac}-MBA)$_{0.5}$ (Fig.~S7). This model reproduces the experimental decoupling while preserving \textit{P2$_1$/m} symmetry and its inversion center. In agreement with previous work,\cite{qian2014quantum, zhang2023spin, joseph2021topological} the monolayer is gapped, whereas the bilayer, the structural unit of 2M-\ce{WS2}, is metallic (Fig.~\ref{fig_dft}a, b). For the \ce{WS2}-\ce{NH3} superlattice, intercalation restores a gapped band structure close to that of the monolayer (Fig.~\ref{fig_dft}c), with molecular states lying well below the Fermi level and contributing negligibly near the gap (Extended Data Fig.~6). The gap persists in the lower-symmetry \ce{WS2}-R-MBA model (\textit{P1}, Fig.~S26), and charge-density analysis shows negligible charge transfer between molecules and layers (Fig.~S27), consistent with experimental characterization. Because the decoupled layers behave as independent two-dimensional sheets, we evaluate the 2D $\mathbb{Z}_2$ invariant of a single layer, which remains nontrivial ($\mathbb{Z}_2 = 1$), indicating that the \ce{WS2}-\ce{NH3} superlattice preserves the quantum spin Hall character of the monolayer (Fig.~\ref{fig_dft}a, c).\cite{qian2014quantum}

These results show that the intercalation-induced interlayer expansion suppresses the interlayer coupling, reopening the gap and restoring the monolayer topological character. The superlattices can thus be viewed as chemically stabilized stacks of electronically decoupled topological layers, mirroring the measured metal-to-insulator transition and the monolayer-like behavior seen in the bulk.

\subsection{Chiral Intercalation}\label{subsec2.5}

Up to this point, the intercalation studied here is essentially non-destructive to the \ce{WS2} host, preserving the local chemical environment and the crystallographic symmetry of the layers, aside from the spatial rearrangements described above. However, beyond acting as electronically inert spacers, chiral intercalants have previously been shown to induce chiral transport in R/S-MBA-intercalated \ce{TaS2},\cite{qian2022chiral, wan2024unconventional} \ce{TiS2},\cite{bian2023chiral} and \ce{MoS2}.\cite{bian2022hybrid} We therefore examine the host-guest interaction more closely, focusing on chiroptical activity induced by chiral intercalation.

\begin{figure}[H]
\centering
\includegraphics[width=1\textwidth]{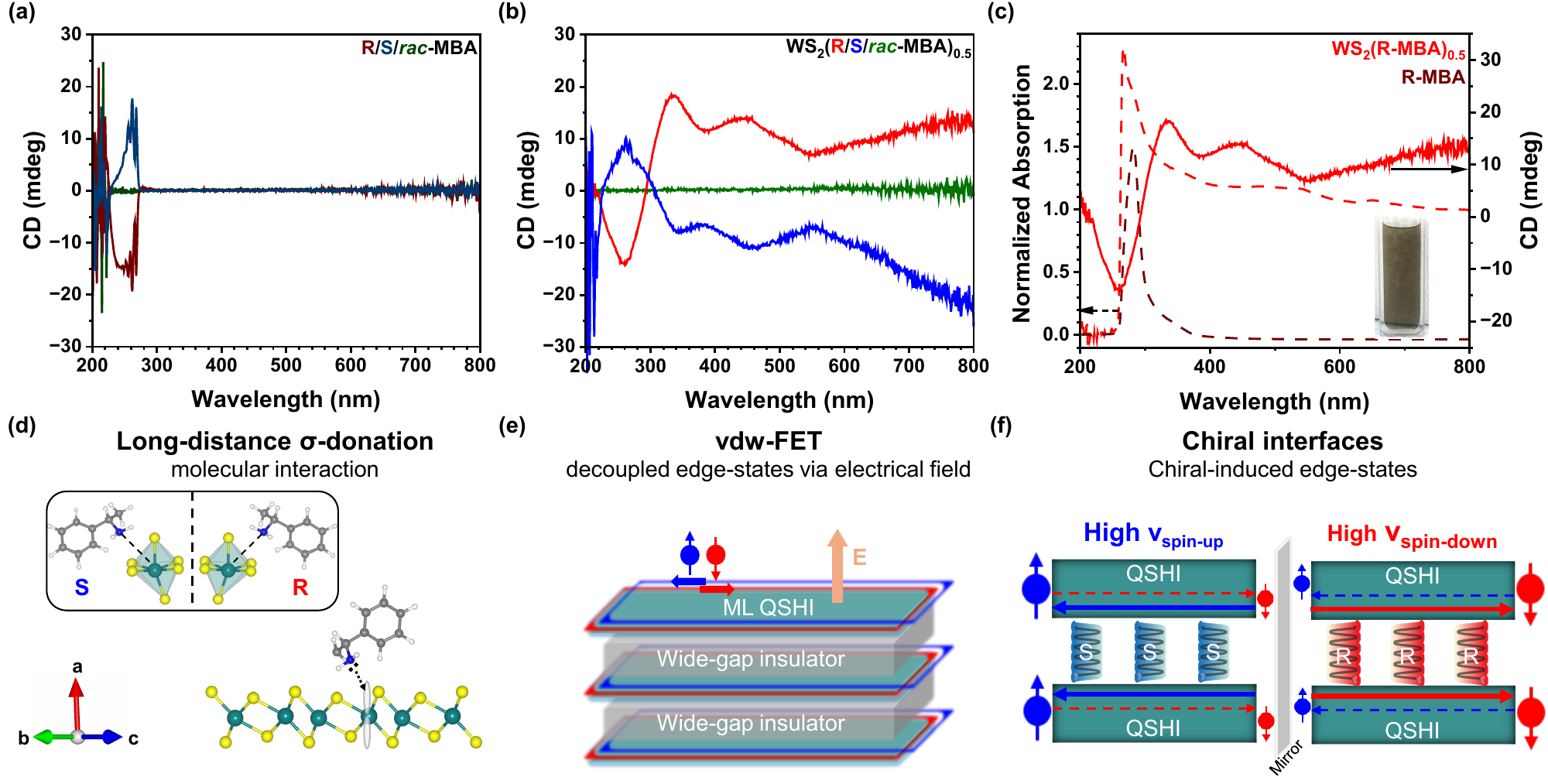}
\caption{\textbf{Optical signatures of chirality transfer in amine-intercalated \ce{WS2}.} CD spectra of R/S/\textit{rac}-MBA molecules (a) and of exfoliated \ce{WS2}(R/S/\textit{rac}-MBA)$_{0.5}$ flakes in IPA (b). Comparison of the optical absorption spectra of R-MBA and \ce{WS2}(R-MBA)$_{0.5}$, together with the corresponding CD response of \ce{WS2}(R-MBA)$_{0.5}$ (c). An exfoliated \ce{WS2} solution used for CD analysis is shown in the inset of panel c. Schematic illustration of the molecular-level interaction between amine and \ce{WS2} (d). Conceptual vdW-FET architecture\cite{qian2014quantum} based on alternating QSHI monolayers and wide-gap insulating layers (e). Proposed chiral interface between chiral amines and \ce{WS2}, enabling decoupling of edge states and tunable spin-channel velocity (f).Supplementary characterization of other amines and amine-intercalated phases is provided in Extended Data Fig.~7.}\label{fig6}
\end{figure}

Free R/S-MBA molecules show circular dichroism (CD) only below 280 nm (Fig.~\ref{fig6}a). In contrast, exfoliated \ce{WS2}(R/S-MBA)$_{0.5}$ dispersions in isopropanol (IPA) exhibit pronounced CD across the entire measured range (Fig.~\ref{fig6}b), closely following the broad absorption features of 1T$'$-\ce{WS2} (Fig.~\ref{fig6}c). The chiral amines therefore imprint a chiroptical response onto the electronic transitions of otherwise achiral \ce{WS2} layers. Similar chiroptical responses imparted by chiral molecules to inorganic layers have also been observed in covalently modified \ce{TiNCl}\cite{zhou2025mxenoids} and amine-intercalated 1T-\ce{TiS2}.\cite{bian2023chiral} Two controls confirm that this response is intrinsic rather than arising from scattering or linear dichroism: (1) the R and S samples produce mirror-image CD spectra (Fig.~\ref{fig6}b), whereas the achiral intercalates, \ce{WS2}(\textit{rac}-MBA/DDA/HexA)$_x$, and pristine 2M-\ce{WS2} show similar absorption but no measurable CD signal (Fig.~\ref{fig6}b, Extended Data Fig. 7d); (2) the alternating positive and negative CD features, together with the offsets between CD and absorption peaks (Extended Data Fig. 7c), resemble those observed in quantum dots capped with chiral ligands,\cite{varga2017cdse} where they have been attributed to chiral-induced splitting of excitonic states into sublevels with opposite angular momentum.\cite{ben2016probing}

Notably, the inorganic layers themselves appear centrosymmetric: in-plane diffraction of \ce{WS2}(R-MBA)$_{0.5}$ is consistent with \textit{P2$_{1}$/m}, which retains an inversion center even after chiral intercalation. Therefore, rather than arising from a chiral \ce{WS2} lattice, the observed chiral optical responses reflect coupling between the \ce{WS2} electronic transitions and the chiral molecular environment, analogous to the induced CD observed in achiral quantum dots capped with chiral ligands, where the atomic structure of the inorganic core remains largely intact.\cite{ben2016probing, varga2017cdse} We describe this coupling as long-distance $\sigma$-donation, a weak molecular interaction mediated by epitaxial N coordination that locally perturbs the \ce{WS2} electronic states without significant atomic displacement (Fig.~\ref{fig6}d). Such a $\sigma$-donation model is consistent with the original proposal for neutral molecular intercalation in 2H-\ce{TaS2},\cite{beal1973intercalation, schollhorn1977demonstration} where donation from the nitrogen lone pair to the metal $d_{z^2}$ orbital was proposed based on small shifts in guest-molecule vibrational frequencies ($<$50 cm$^{-1}$),\cite{tofield1977vibration, rao1979intercalation} characteristic absorption spectra,\cite{beal1973charge} and stoichiometric host-guest molecular ratios.\cite{gamble1971some} The minimal net charge transfer and the preserved \ce{WS2} framework observed in our system are inconsistent with later interpretations invoking redox chemistry associated with water or impurity incorporation.\cite{schollhorn1979ionic, johnson1980amine, mckelvy1987intercalation} Overall, the results of this work support weak $\sigma$-donation as the dominant host-guest interaction, locally modifying the electronic states of \ce{WS2} while preserving the overall lattice structure.

Beyond the local chemistry, these superlattices, built from alternating topological and insulating layers, realize within a single crystal the vdW topological field-effect-transistor (vdW-TFET) heterostructure proposed earlier (Fig.~\ref{fig6}e).\cite{qian2014quantum} Given the coexistence of chiral polarization and the 2D TI state in 1T$'$-\ce{WS2}, a natural question is whether chirality can affect the spin-channel velocity or population of the QSH state. The abundant interfaces between inorganic layers and chiral spacers are well suited to proximity effects from chiral-induced spin selectivity (CISS)\cite{evers2022theory, wolf2022unusual, liu2025precision} or Rashba spin-orbit coupling (Fig.~\ref{fig6}f).\cite{ortiz2016generic, jana2020organic} Finally, while mechanical exfoliation of bulk 2M-\ce{WS2} is hindered by strong interlayer interactions, intercalation weakens these interactions and facilitates exfoliation: preliminary experiments already yield few-micrometer flakes down to monolayers, opening the way to low-dimensional transport studies of the interplay between topology and chirality (Extended Data Fig. 8).

\section{Conclusion}\label{sec3}

In conclusion, we establish neutral-molecule intercalation of 2M-\ce{WS2} as a wet-chemical route to electronically decoupled, monolayer-like 1T$'$-\ce{WS2} superlattices. The intercalation drives pronounced interlayer expansion and ordered layer shifts while preserving the intrinsic 1T$'$-\ce{WS2} framework. In contrast to the metallic, superconducting parent 2M phase, the resulting superlattices are insulating and, according to DFT, retain the topological character expected for the monolayer, thereby providing access to monolayer-like electronic regimes within bulk crystals.

Beyond serving as a platform for exploring the electronic behavior of monolayers, these superlattices constitute vdW heterostructures within single crystals, providing an alternative to mechanically assembled stacks of alternating active and insulating layers. The host-guest interaction introduces an additional degree of freedom for tuning the electronic structure through interfacial proximity effects, as demonstrated by the chirality imprinting. More generally, molecular intercalation of vdW crystals offers a distinct chemical route to engineering emergent electronic functionality in layered quantum materials.

\section{Methods}\label{sec4}
\subsection{Synthesis}\label{subsec4.1}
2M-\ce{WS2} crystals were synthesized by acid treatment of \ce{K_{x}WS2} according to previously reported procedures.\cite{song2023acid} The obtained crystals were isolated from the acid matrix, washed with water and methanol, and dried under vacuum. All amines were purchased from Oakwood Chemical and stored over activated 4~\AA\ molecular sieves to minimize residual water content. All synthetic procedures were carried out in an Ar-filled glovebox, and all glassware, including vials and pipettes, was oven-dried at 120~$^\circ$C for at least 48 h prior to transfer into the glovebox. Although the isolated 2M-\ce{WS2} crystals are chemically robust and air-stable, stringent precautions were taken to minimize exposure to water, oxygen, and other impurities to ensure well-controlled reactions with the amines.

A typical intercalation procedure for all amines is described below. First, 20 mg of 2M-\ce{WS2} crystals in a 24 mL vial were rinsed with 1 mL of the corresponding amine, and the washing solution was discarded. Subsequently, 4 mL of fresh amine was added, and the capped vial was sealed with electrical tape and heated at 65~$^\circ$C for 7 days using a heating block on a temperature-controlled hot plate. The reaction solution was then removed and replaced with 3 mL of fresh amine, followed by resealing and heating at 100~$^\circ$C for an additional 7 days. After completion of the reaction, the resulting crystals were isolated from the liquid and dried under vacuum at room temperature until no visible residual solvent remained.

Except for HexA, all intercalation reactions yielded fully intercalated \ce{WS2}(amine)$_{0.5}$ phases, in which the interlayer separation corresponds to approximately twice the molecular length. Under the same reaction conditions, \ce{WS2}(HexA)$_{0.3}$ was also intercalated but showed an interlayer spacing corresponding to approximately one molecular length (Fig.~S20). Detailed studies of the intercalation mechanism and controlled access to intermediate phases will be presented in future work. Partial intercalation of \ce{WS2}(R-MBA)$_{x}$ was achieved by heating at 65~$^\circ$C for 7 days only, yielding \ce{WS2}(R-MBA)$_{0.34}$. All intercalated materials were generally air-stable, although prolonged exposure to ambient moisture could lead to mild water uptake by the intercalated amines and additional interlayer expansion.

For de-intercalation, 20 mg of \ce{WS2}(R-MBA)$_{0.5}$ crystals were immersed in 5 mL of 0.02 M \ce{HNO3} solution and shaken gently at 300 rpm for 12 h using a Scilogex SK-0330-Pro shaker. The resulting de-intercalated crystals were isolated, washed with water and methanol, and dried in air prior to characterization. Partial de-intercalation was achieved by limiting the acid treatment time to 2 h.

\subsection{Characterization}\label{subsec4.2}
PXRD patterns were obtained with 0.5 mm borosilicate glass capillaries on a STOE STADI P powder diffractometer with Mo $K_{\alpha1}$ radiation and a Dectris Mythen 2R 1K detector, in the $2\theta$ range from 1 to $45^\circ$. PXRD patterns used for Pawley and Rietveld refinements were collected over 12 h. Structural refinements were performed using TOPAS V7 (Bruker AXS), and structural models were constructed with BIOVIA Materials Studio 2017 software and visualized in VESTA.

K, W, and S ratios were determined using an Agilent 5800 ICP-OES instrument. Samples were first treated with trace-metal-grade \ce{HNO3}/HCl (9:1) in a MicroDigest system at 80~$^\circ$C for 5 min, resulting in the formation of white-yellow \ce{WO_x} precipitates. The mixtures were subsequently neutralized with aqueous \ce{NH3.H2O} to pH 8 to 9, under which conditions the precipitates gradually dissolved over time. The resulting solutions were filtered through hydrophilic syringe filters and diluted with Milli-Q water prior to analysis.

XPS measurements were performed on a ThermoFisher K-Alpha spectrometer. Samples were transferred directly from the glovebox into the XPS chamber using a vacuum-transfer module. Spectra were analyzed and fitted using CasaXPS and referenced to the adventitious C 1s peak at 284.5 eV.

Raman spectra were collected using a Horiba Raman spectrometer equipped with a 532 nm laser and a neutral-density filter (ND = 0.6).

Fourier-transform infrared (FTIR) spectra were acquired in reflection mode on a Thermo Nicolet 6700 FTIR spectrometer.

SEM images and EDS data were collected using a Quanta 200 FEG environmental SEM equipped with an Oxford energy-dispersive X-ray spectrometer.

Cross-sectional TEM lamellae were cut out of bulk single crystals using a Helios DualBeam focused ion beam (FIB)/SEM system and thinned to approximately 80~nm (Fig.~S18). The lamellae were transferred to Mo lift-out TEM grids and immediately loaded into the high-vacuum chamber of the S/TEM microscope to minimize air exposure. Plan-view TEM samples were prepared by Scotch-tape exfoliation and transferred directly onto Si$_3$N$_4$ TEM grids. SAED and HAADF-STEM images were collected using a Titan Cubed Themis 300 double-Cs-corrected S/TEM operated at 300 kV, providing a spatial resolution of 0.07 nm and an energy resolution of 0.8 eV. The microscope is additionally equipped with a Super-X EDS system for elemental mapping and a Gatan Quantum SE/963 P post-column energy filter for energy-filtered TEM measurements.

DSC measurements were performed using a PerkinElmer HyperDSC DSC8500 calorimeter. Heating rates of 10~$^\circ$C min$^{-1}$ were used for \ce{WS2}(MBA)$_{0.5}$ samples and 20~$^\circ$C min$^{-1}$ for \ce{WS2}(Hex)$_{0.3}$  and \ce{WS2}(DDA)$_{0.5}$ samples. Cooling rates were fixed at 20~$^\circ$C min$^{-1}$.

TGA measurements were conducted under flowing nitrogen using a PerkinElmer TGA-8000 thermogravimetric analyzer.

Four-probe electrical transport measurements were carried out on a Physical Property Measurement System (PPMS, Quantum Design) using the ETO option. After mechanical cleavage with adhesive tape (Scotch Magic tape), crystals were fixed to sapphire substrates with GE varnish and contacted using four gold wires and silver paint (Fig. S22).

DC magnetic measurements were performed on an MPMS-3 (Quantum Design) operating in vibrating-sample magnetometer (VSM) mode. Pressed pellets of crystals were used for most characterizations, while anisotropic measurements on single crystals were additionally performed (Fig. S25). To minimize remanent magnetic fields in the SQUID magnet, the magnetic field was oscillated to 0 T following the standard Quantum Design demagnetization procedure.

Ultraviolet-visible-near-infrared (UV-vis-NIR) absorption spectra were collected using a Varian Cary 5000 spectrophotometer, and CD spectra were measured using a Chirascan CD spectrometer (Applied Photophysics). For sample preparation, approximately 5 mg of material was dispersed in 1 mL of IPA and sonicated for 10 min to obtain homogeneous black exfoliated dispersions (Fig.~S2).  Amine solutions were prepared by dissolving 10 $\mu$L of amine in 5 mL of IPA. Measurements were performed using quartz cuvettes.

Atomic force microscopy (AFM) topography and Kelvin probe force microscopy (KPFM) measurements were performed using a Park NX-Hivac system under a nitrogen atmosphere at a pressure of $\sim100$ mbar. A conductive Pt/Ir-coated PPP-EFM cantilever (resonance frequency about 75 kHz, spring constant 3 N m$^{-1}$) was used for all measurements. AFM topography was acquired in contact mode with a normal load of 50 nN. KPFM measurements were performed in sideband KPFM mode with an AC bias voltage of 1.5 V applied at a modulation frequency of 1.5 to 3 kHz. Mechanical exfoliation was carried out using adhesive tape (Scotch Magic tape) in an Ar-filled glovebox.

\subsection{Calculations}\label{subsec4.3}
DFT calculations were performed on proposed crystal structures based on experimental data using Vienna ab initio simulation package (VASP) 6.4.2. \cite{Kresse1993AbMetals, Kresse1994AbGermanium, Kresse1996EfficiencySetb, Kresse1996EfficientSet} The Perdew-Burke-Ernzerhof (PBE) generalized gradient approximation (GGA)\cite{Perdew1996GeneralizedSimplec} functional was used for exchange-correlation functional, with the recommended potpaw-PBE.64 projector-augmented wave (PAW) pseudopotentials provided by VASP.\cite{Blochl1994ProjectorMethod, Kresse_ultrasoft_1999} The calculations used a plane-wave energy cutoff of 550 eV, a $\Gamma$-centered $1\times8\times8$ Monkhorst-Pack $\mathbf{k}$-point grid, \cite{Monkhorst_kpoints_1976} and included spin-orbit coupling (SOC); the electronic self-consistency criterion was $10^{-6}$ eV. Band structures and density of states were calculated using the \texttt{sumo} Python package. \cite{sumo} Band structures were computed with full paths (Extended Data Fig. 6) as well as the selected in-plane path connecting $\Gamma = (0,0,0)$ and Y$= (0.5,0.5,0)$ for better comparison with previous literature results.\cite{qian2014quantum, zhang2023spin}. 

Topological classification was performed within the framework of topological quantum chemistry.\cite{Bradlyn2017TopologicalChemistry} Irreducible representations of the occupied bands at the time-reversal-invariant momenta were extracted from the VASP wavefunctions with \textit{vasp2trace} and analyzed with the \textit{Check Topological Mat.} tool of the Bilbao Crystallographic Server, \cite{Vergniory2019CompleteCatalogue, Vergniory2021AllTopological, Aroyo2006Bilbao} yielding the symmetry-indicator invariants, $\mathbb{Z}_2$.

\backmatter

\bmhead{Supplementary information}
The supporting information is available free of charge at (PDF).

\bmhead{Acknowledgements}
This research was supported by the Gordon and Betty Moore Foundation's EPIQS initiative (grant number GBMF9064), the Princeton Center for Complex Materials, a National Science Foundation (NSF)-MRSEC program (DMR-2011750), the Princeton Catalysis Initiative (PCI), and the Packard Foundation. The authors acknowledge the use of Princeton's Imaging and Analysis Center, which is partially supported by the Princeton Center for Complex Materials, a NSF-MRSEC program (DMR-2011750). The authors acknowledge the use of the biophysics core facility of the Department of Chemistry, Princeton University. The simulations presented in this work were performed on computational resources managed and supported by Princeton University's Research Computing. J.C. acknowledges the support of the NSF under Grant No. DMR-1942447 and from the Flatiron Institute, a division of the Simons Foundation. The authors thank Mohamed Nabil Yacine Lhachemi for valuable discussions on topological calculations.

\section*{Declarations}

J.X. designed and performed the experiments and analyzed the data. F.K. performed the computational studies. F.Y., G.C., and X.S. performed the HRSTEM studies. J.M.M. performed transport measurements, and C.J.P. performed Raman spectroscopy measurements on the crystals. N.R., Y.B., and M.B.S. carried out mechanical exfoliation and related characterization of nanoflakes. J.C. contributed to discussions of the theoretical results and underlying physical concepts. L.M.S. supervised the project, and J.X. and L.M.S. co-wrote the manuscript. All authors discussed the results and commented on the manuscript.

\bibliography{Reference}

\begin{appendices}

\renewcommand{\thefigure}{\arabic{figure}}
\renewcommand{\figurename}{Extended Data Fig.}

\section{Extended Data}\label{secExtended data}
\begin{figure}[h]
\centering
\includegraphics[width=1\textwidth]{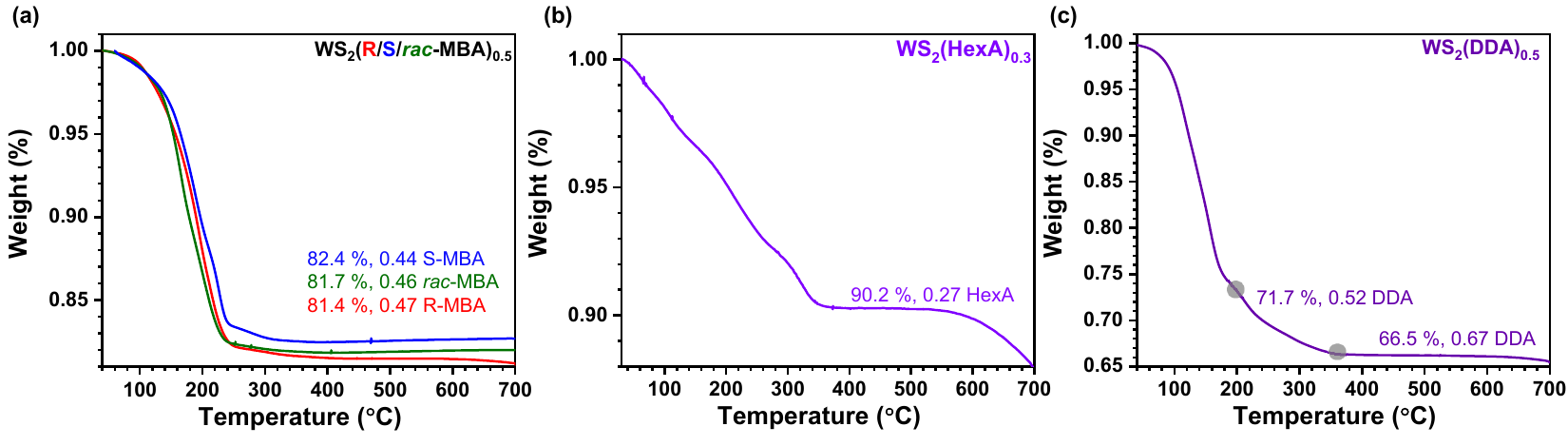}
\caption{\textbf{Thermogravimetric analysis of the amine intercalates.} TGA curves of \ce{WS2}(MBA)$_{0.5}$ (a), \ce{WS2}(HexA)$_{0.3}$ (b), and \ce{WS2}(DDA)$_{0.5}$ (c). Because DDA is solid at room temperature, its removal after reaction is less efficient, giving a higher residual content than 0.5, which is expected for a fully intercalated phase. The nominal composition of \ce{WS2}(DDA)$_{0.5}$ is estimated from the first mass-loss step in the TGA profile.}\label{ED-TGA}
\end{figure}

\begin{figure}[h]
\centering
\includegraphics[width=0.8\textwidth]{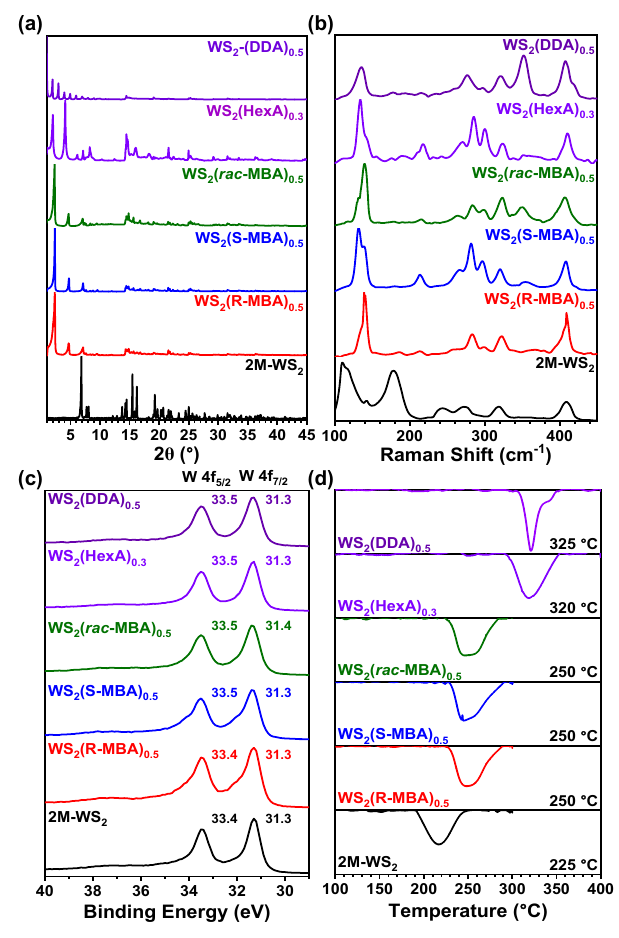}
\caption{\textbf{Chemical characterization of all amine intercalates.} PXRD (a), Raman (b), XPS (c), and DSC (d) data of \ce{WS2}(amines)$_{x}$. The strongest PXRD reflection of \ce{WS2}(Hex)$_{0.3}$ appears at 4.12$^\circ$ (d = 9.9~\AA), rather than the low-angle peak at 2.02$^\circ$ (d = 20~\AA), indicating predominantly single-layer HexA intercalation with a minor contribution from a double-layer phase. This is consistent with the intermediate composition (x = 0.3 relative to 0.5) and the cross-sectional HRSTEM analysis (Fig. S20). For \ce{WS2}(DDA)$_{0.5}$, the (100) reflection lies below the instrumental detection limit of 1$^\circ$, while the (200) peak at 1.99$^\circ$ (d = 20.5~\AA) indicates an interlayer expansion of ~40~\AA\, consistent with double-molecular-layer intercalation.}\label{ED-Chem}
\end{figure}

\begin{figure}[h]
\centering
\includegraphics[width=0.9\textwidth]{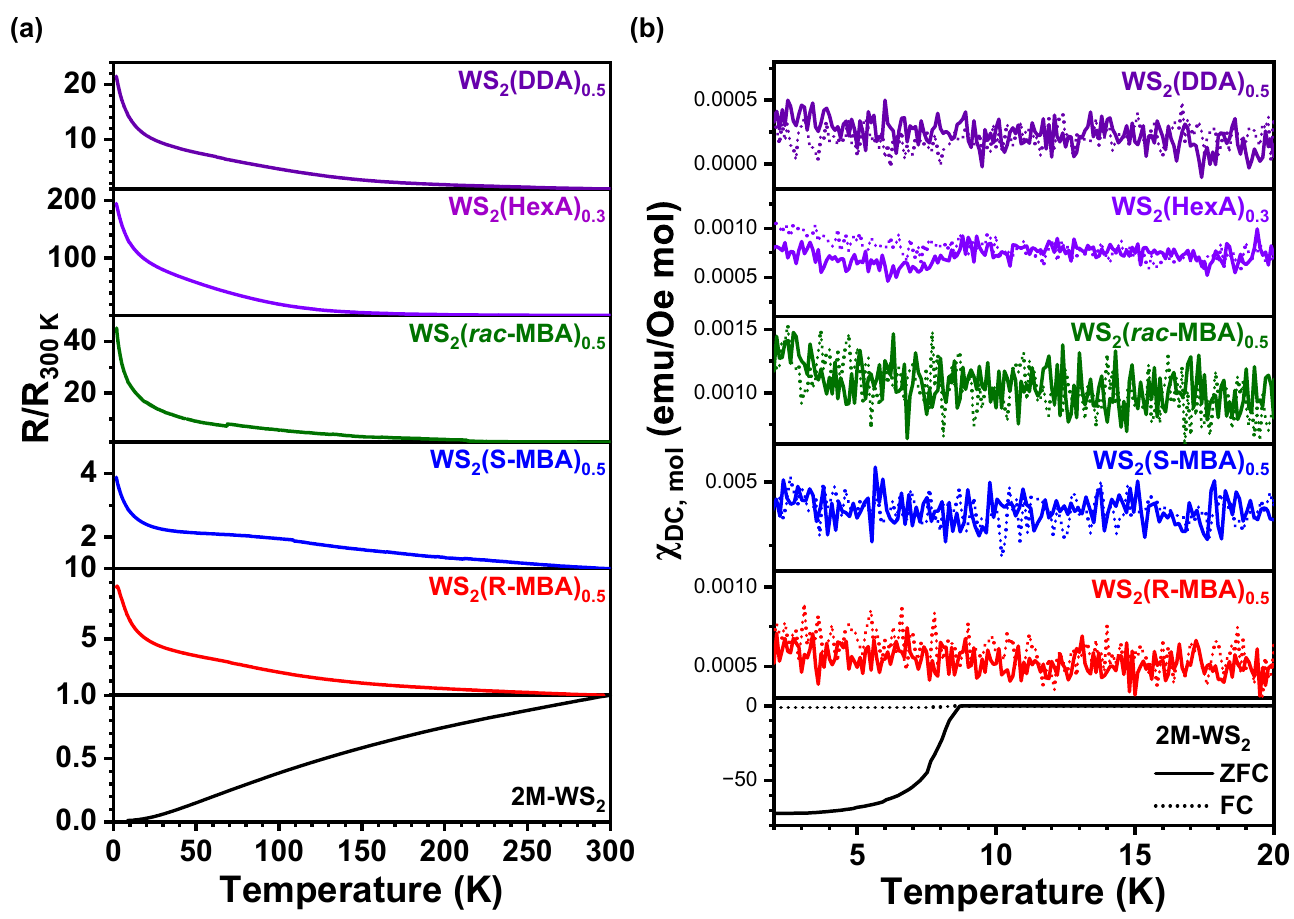}
\caption{\textbf{Transport and magnetic properties of all amine intercalates.} Temperature-dependent resistance (a) and magnetic susceptibility (b) of \ce{WS2}(amines)$_{x}$. Electrical resistance was measured on single crystals using a four-probe configuration (Fig. S22), while magnetization was measured on powdered samples under zero-field-cooled (ZFC) and field-cooled (FC) conditions with an applied field of 5 Oe. Corresponding crystal-based magnetic measurements are shown in Fig. S24.}\label{ED-Phys}
\end{figure}

\begin{figure}[h]
\centering
\includegraphics[width=1\textwidth]{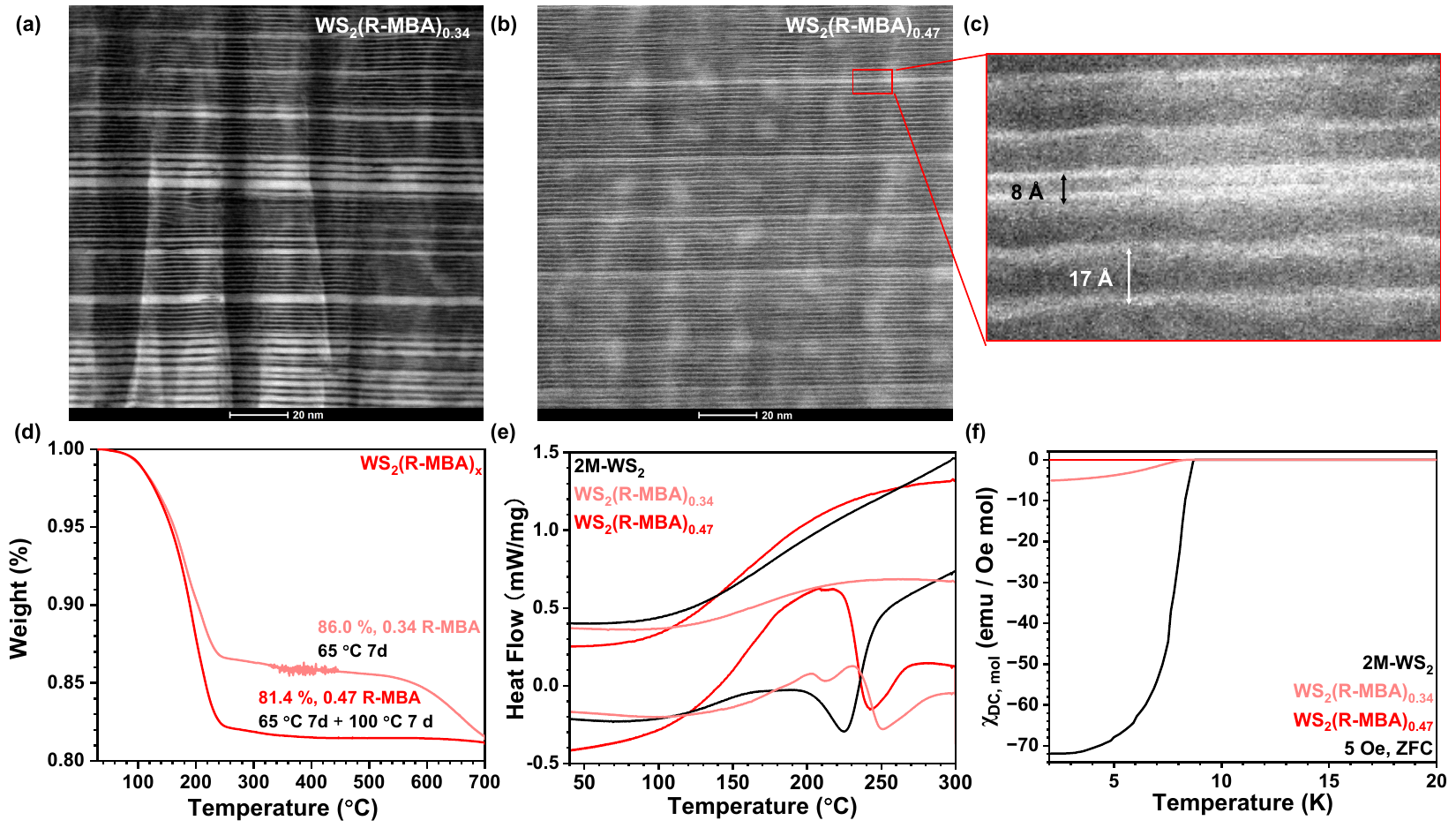}
\caption{\textbf{Structural and physical evolution during intercalation.} Cross-sectional HRSTEM images of partially intercalated \ce{WS2}(R-MBA)$_{0.34}$ (a) and fully intercalated \ce{WS2}(R-MBA)$_{0.47}$ (b, with a zoom-in view in panel c). Corresponding TGA (d), DSC (e), and magnetization data (f) are also shown. HRSTEM analysis reveals that non-intercalated 2M domains remain in \ce{WS2}(R-MBA)$_{0.34}$. DSC suggests a minor 2M-to-2H phase transformation in addition to the dominant intercalated 1T$'$-to-2H transition. Magnetization indicates that the residual 2M phase contributes a comparable superconducting volume fraction. For \ce{WS2}(R-MBA)$_{0.47}$, HRSTEM images reveal predominantly fully intercalated regions with double-molecular expansion, together with minor domains showing single-molecular expansion, consistent with an overall composition of 0.47 rather than 0.5. Although regions with single-molecular-layer expansion are present, no residual \ce{WS2} stacking domains or pristine 2M phase are observed in the TEM images, nor are they evidenced by DSC or superconductivity measurements. Additionally, minor domains of single-molecular expansion may originate from partial deintercalation during TEM sample preparation, where FIB cutting may locally generate sufficient heat. }\label{ED-Partial In}
\end{figure}

\begin{figure}[h]
\centering
\includegraphics[width=0.9\textwidth]{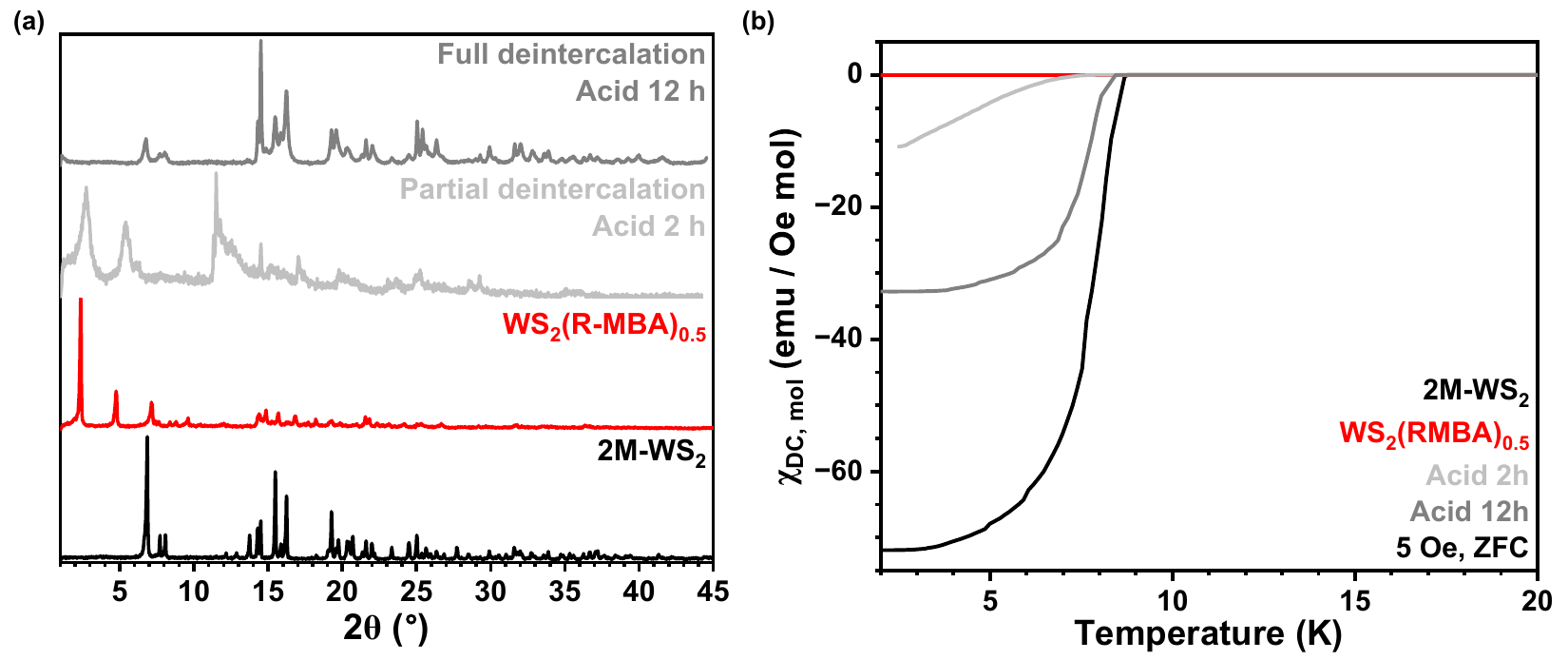}
\caption{\textbf{Structural and magnetic evolution during de-intercalation.} Evolution of structure and properties during de-intercalation, monitored by PXRD (a) and magnetization (b). The partially de-intercalated phase, which contains a minor 2M component, still shows clear superconductivity, and the superconducting fraction increases and becomes more volumetrically extended upon full de-intercalation.}\label{ED-Partial De-in}
\end{figure}

\begin{figure}[h]
\centering
\includegraphics[width=1.0\textwidth]{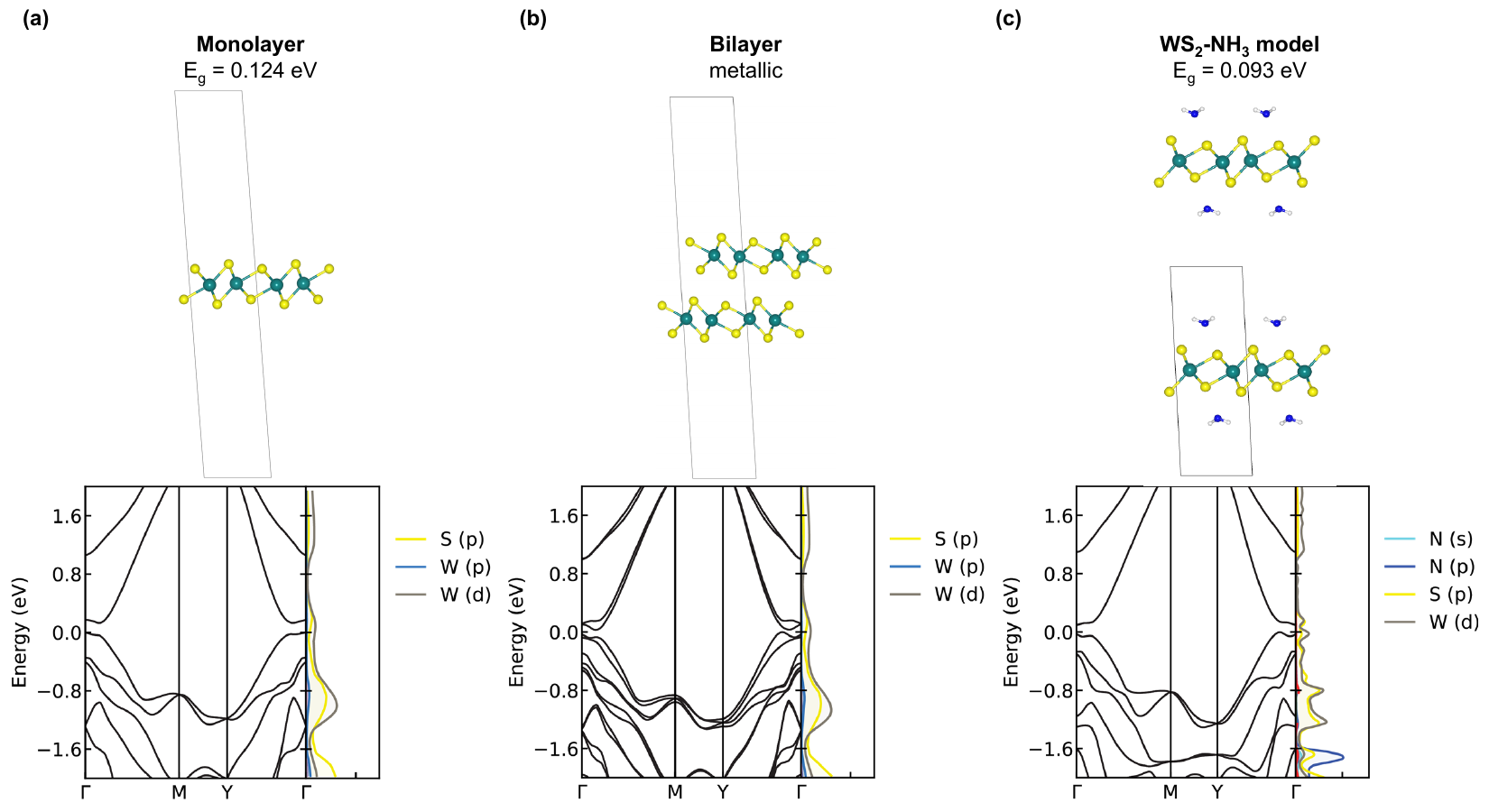}
\caption{\textbf{Band structures along the full Brillouin-zone paths.} DFT-calculated electronic band structures for 1T$'$-\ce{WS2} monolayer (a), bilayer (b), and the \ce{WS2}-\ce{NH3} model (c). The structure of \ce{WS2}-\ce{NH3} was derived from Rietveld refinement of the PXRD pattern of \ce{WS2}(\textit{rac}-MBA)$_{0.5}$ (Fig. S7). The monolayer exhibits an indirect band gap of 0.124 eV and a direct band gap of 0.149 eV. For the \ce{WS2}-\ce{NH3} superlattice, the corresponding indirect and direct band gaps are 0.093 eV and 0.122 eV, respectively.}\label{Band-full}
\end{figure}

\begin{figure}[h]
\centering
\includegraphics[width=1\textwidth]{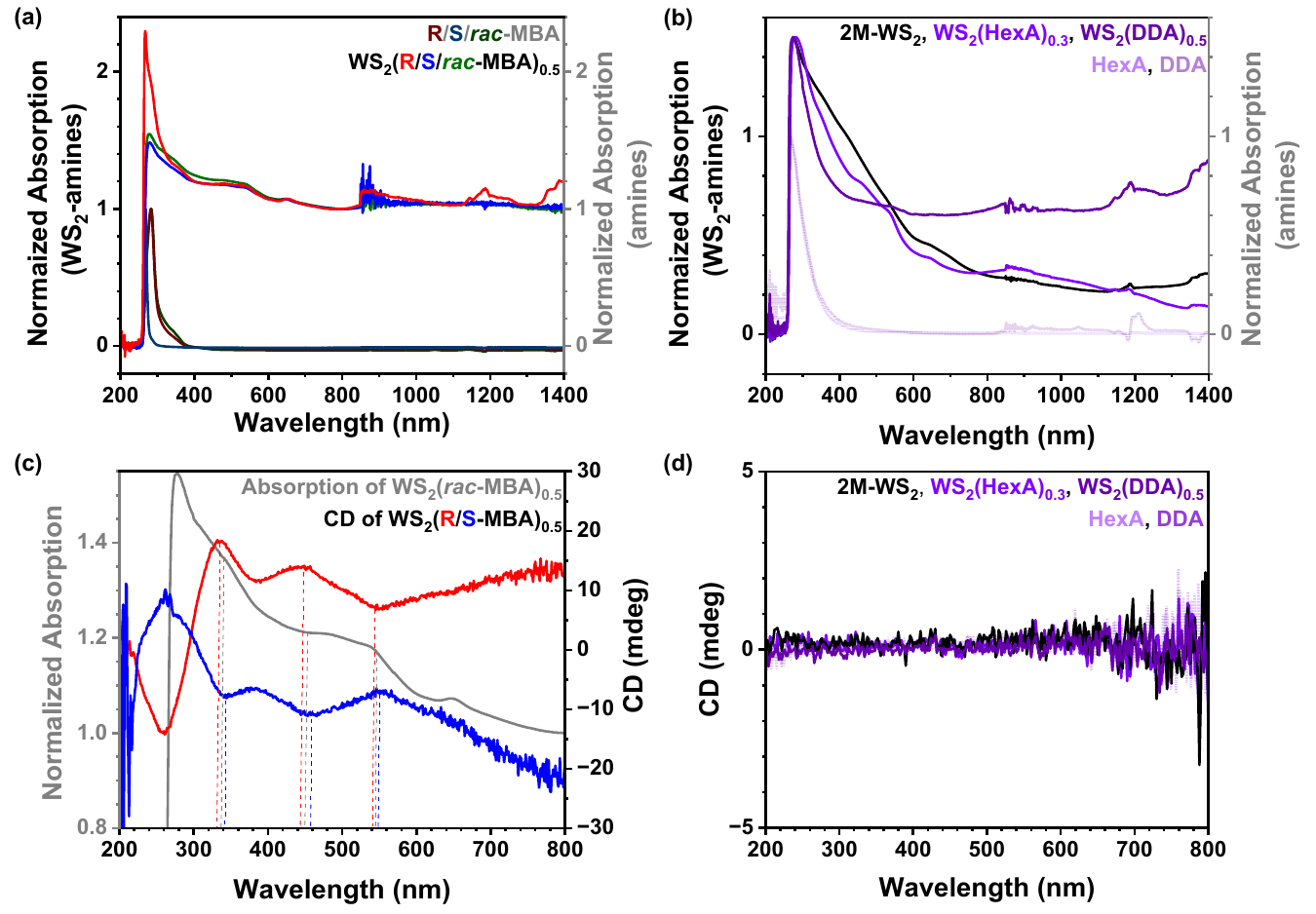}
\caption{\textbf{Optical characterization of all amine intercalates.} UV-vis-NIR absorption spectra of \ce{WS2}(amines)$_{x}$ and the corresponding free amines (a, b). Comparison of absorption and CD spectra of \ce{WS2}(R/S-MBA)$_{0.5}$ (c) and CD spectra of the non-chiral \ce{WS2}(amines)$_{x}$ (d). In panels a and b, the noisy and discontinuous spectral region around 900 nm is attributed to instrumental noise arising from light-source switching. The exfoliated dispersions of intercalated phases cannot be maintained at identical concentrations due to an uncontrollable sonication process, which is further complicated by various crystal sizes and different exfoliation efficiencies among intercalate species. Therefore, the optical absorption spectra were normalized for comparison. However, the signal-to-noise ratio varies among samples, and the spectral profiles also show some variations. Regardless, all \ce{WS2}-based materials exhibit similar band-like absorption features with identical sub-band peak positions.}\label{ED-Opt}
\end{figure}

\begin{figure}[h]
\centering
\includegraphics[width=0.8\textwidth]{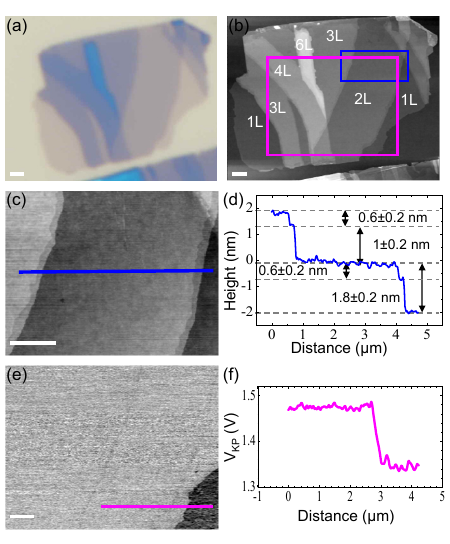}
\caption{\textbf{Mechanical exfoliation to the atomic limit.} Mechanical exfoliation of \ce{WS2}(R-MBA)$_{0.5}$ into the atomic limit. Optical microscopy image (a) and AFM topographic image (b) of an exfoliated flake, showing regions with thicknesses ranging from 1L to 6L. The blue and purple highlighted regions correspond to panels c,d and e,f, respectively. High-resolution AFM topography of the blue selected region (c) and the corresponding height profile extracted along the blue line (d). Kelvin probe force microscopy (KPFM) surface potential map (e) and the corresponding surface potential profile extracted along the magenta line (f). Scale bars in all images correspond to 1 $\mu$m. The AFM measurements reveal a predominant step height of 1.8 $\pm$ 0.2 nm, which can be resolved into two substeps: a 0.6 $\pm$ 0.2 nm step assigned to a single \ce{WS2} layer and a 1.0 $\pm$ 0.2 nm step attributed to an intercalated molecular layer. KPFM measurements show negligible variation in surface potential between regions of different \ce{WS2} thickness, while a potential difference of approximately 130 mV is observed between the \ce{SiO2} substrate and the monolayer \ce{WS2} region. The nearly thickness-independent surface potential indicates that the exposed surfaces are predominantly terminated by \ce{WS2} layers rather than intercalant molecules, consistent with molecular species residing within the interlayer galleries. This surface termination leaves large domains of exposed 1T$'$-\ce{WS2} and is advantageous for probing the intrinsic properties of exfoliated monolayer 1T$'$-\ce{WS2}.}\label{ED-Exfoilation}
\end{figure}

\end{appendices}

\end{document}